\documentclass[10pt,onecolumn]{IEEEtran}
\usepackage{amsfonts,amsmath,amssymb,amsthm,caption,algorithmic}
\usepackage{amsbsy}
\usepackage{graphicx,multirow,bm}
\usepackage{color}
\usepackage[para]{threeparttable}

\makeatletter
\newtheoremstyle{mythm}{3pt}{3pt}{}{16pt}{\bfseries}{:}{.5em}{}
\theoremstyle{mythm}

\newtheorem{construction}{Construction}
\newtheorem{theorem}{Theorem}

\newtheorem{definition}{Definition}
\newtheorem{Example}{Example}

\newtheorem{lemma}{Lemma}

\newcommand{\cC}{\mathcal{C}}

\newcommand{\cE}{\mathcal{E}}
\newcommand{\cF}{\cE}

\newcommand{\cR}{\mathcal{R}}

\newcommand{\Z}{\mathbb{Z}}
\newcommand{\F}{\mathbb{F}}

\begin{document}

\title{A Class of Minimum Storage Cooperative Regenerating Codes with Low Access Property
\author{ Xing Lin, Han Cai,~\IEEEmembership{Member,~IEEE}, and Xiaohu Tang,~\IEEEmembership{Senior Member,~IEEE}}
\thanks{X. Lin, H. Cai, and X. H. Tang are with the Information Coding \& Transmission Key Lab of Sichuan Province, CSNMT Int. Coop. Res. Centre (MoST), Southwest Jiaotong University, Chengdu, 611756, China (e-mail: 505976471@qq.com, hancai@swjtu.edu.cn, xhutang@swjtu.edu.cn).}
}
\date{}
\maketitle

\begin{abstract}
In this paper, a new repair scheme for a modified construction of MDS codes is
studied. The obtained repair scheme has optimal bandwidth for multiple failed
nodes under the cooperative repair model. In addition, the repair scheme has relatively low access property, where
the number of data accessed is less than two times the optimal value.
\end{abstract}


\section{Introduction}\label{I}

\IEEEPARstart{I}n large-scale cloud storage and distributed file systems, such as Amazon
Elastic Block Store (EBS) and Google File System (GoogleFS), one direct result of the massive scale
is that node failures are the norm and not the exception.
Usually, to ensure high reliability, the simplest solution
is a straightforward replication of data packets across different nodes, for example store three
copies.  However, this duplication solution suffers from a large storage overhead.
An alternative solution to protect the data from node failures is via erasure coding.
For example, compared with replication, $[n,k]$ maximum distance separable (MDS) codes are capable to provide a dramatic improvement in redundancy, which encode $k$ information symbols
to $n$ symbols and store them across $n$ nodes. However, MDS codes suffer a reconstruction problem, i.e.,
when one node fails, the system recovers it at the cost of contacting
$k$ surviving nodes (helper nodes) and downloading whole data size of the original file.

Two important indicators to measure the repair efficiency are the amount of data transmitted in the repair, called repair bandwidth, and the amount of data accessed at the helper nodes, which are used to measure the network usage and disk I/O cost, respectively.
In the seminal paper \cite{dimakis2010network}, Dimakis \textit{et al.} proposed regenerating codes, which can be seen as vector MDS codes attaining  the tradeoff between the storage overhead and repair bandwidth for repairing single node.
Among regenerating codes, the minimum storage regenerating (MSR) codes is the most attractive due to its storage efficiency.  Especially, the MSR codes is called optimal-access MSR codes if the amount of data accessed during the repair is equal to repair bandwidth.  Later on, numerous constructions of MSR codes were deeply studied by \cite{han2015update}, \cite{Han15update}, \cite{Li15Frame}--\cite{rashmi2011optimal} and \cite{tamo2012zigzag}--\cite{ye2017explicit}, where the optimal-access MSR codes were shown in \cite{li2018generic}, \cite{tamo2012zigzag}--\cite{Wang11long} and \cite{ye2017explicit}. Readers can refer to \cite{balaji2018overview} and  \cite{dimakis2011survey} for more MSR code constructions.

Regenerating codes usually dispose of the failure of a single node, however, the failure of multiple nodes is common in large-scale storage system \cite{ford2010availability}. For multi-nodes failure, there are two repair models. One is the centralized repair model, in which a special node (data center) repairs all failed nodes.
The other is the cooperative repair model, in which the failed nodes are regenerated in a distributed and cooperative manner.
The cut-set bounds that give a tradeoff between the storage overhead and repair bandwidth for two repair models were derived in \cite{cadambe2010centralized} and \cite{hu2010cooperative}, \cite{shum2013cooperative}, respectively. In
\cite{ye2018cooperative}, it is proved that the MDS code attaining the optimal repair bandwidth under the cooperative repair model must also achieve the optimal repair bandwidth under the centralized repair model.
Furthermore, owing to the distributed pattern, the cooperative repair model is more consistent with distributed storage systems than the centralized repair model \cite{zhang2019scalar}.
So, this paper is devoted to the minimum storage MDS code under the cooperative repair model, i.e., the minimum storage cooperative regenerating (MSCR) codes.

In general, the motivation of studying MSCR codes is to provide
explicit construction with small sub-packetization level and good access property,
where the sub-packetization level $N$ represents the size of data stored in each node for MSCR codes.
Let $n$, $k$, $d$, $h$ denote the length, dimension, number of helper nodes, and
the number of failed nodes for an MSCR code, respectively.
Via the product matrix approach, \cite{zhang2019scalar} gave an explicit construction of
MSCR codes
for $d\geq \max\{2k-1-h,k\}$ with the sub-packetization level $N=d-k+h$ for the low code rate cases.
In \cite{ye2018cooperative}, an explicit construction of MSCR codes is provided for all possible values of $n,k,h,d$, i.e., $2\leq h\leq n-d\leq n-k-1$, where the sub-packetization level of the construction is $N=((d-k+h)(d-k)^{h-1})^{\binom{n}{h}}=\exp(\Theta(n^h))$, where $n-k$ is regarded as a constant.
Based on space sharing  Hadamard MSR codes in \cite{ye2017explicit} and
a new repair scheme, \cite{ye2020new} introduced an explicit construction
of MSCR codes for all admissible $n,k,h,d$, where the sub-packetization level is
$N=(d-k+h)\cdot(d-k+1)^n=\exp(O(n))$. %The above  MSCR codes \cite{ye2018cooperative}, \cite{43}, \cite{45} require all data stored in the helper nodes to be access during the process of repair and the field size $|F|\geq (d-k+1)n$.
It should be noted that the MSCR codes in \cite{ye2018cooperative}, \cite{ye2020new} and \cite{zhang2019scalar} require all stored data in the helper nodes to be accessed during repair. In  \cite{zhang2020explicit}, an explicit construction of
 MSCR codes with optimal-access property for all possible $n,k,h,d$ was proposed, where the sub-packetization level
  is $N=(d-k+h)^{\binom{n}{h}}=\exp(\Theta(n^h))$.

The main contribution of this paper is to present a new repair scheme for a class of MSCR code
built on the MSR code in \cite{chen2019explicit},
which is capable to repair multiple failed nodes with optimal repair bandwidth
and low access property. Actually, only the optimal repair
process for a single failed node is known in \cite{chen2019explicit}. In this paper,  the original construction is modified by space sharing technique so that we can find a new
repair scheme for multiple failed nodes with optimal repair bandwidth under cooperative repair model. In particular, the optimal repair scheme has low access
property,  precisely the total amount of data accessed in the repair process is less than two times of the optimal value. As a comparison, in
Table \ref{comparison-1}, we list parameters of the MSCR code together with
known ones, where all the MSCR codes can take all possible values of
$n,k,h,d$, i.e., $2\leq h\leq n-d\leq n-k-1$.

\begin{table}[htbp]
\begin{center}
\caption{Comparisons with the existing MSCR codes that permit
all possible parameters $n,k,h,d$.}\label{comparison-1}
\begin{tabular}{|c|c|c|c|c|}
\hline
&Sub-packetization&\multirow{2}{*}{Field size $|\F|$} &The total amount of data & \multirow{2}{*}{Remark} \\
&level $N$ &  &  accessed in the repair &  \\
\hline\hline
The MSCR code in \cite{ye2018cooperative}&$((d-k+h)(d-k)^{h-1})^{\binom{n}{h}}$&
$|\F|\geq (d-k+1)n$&$dN$& Full access\\
\hline
The MSCR code in \cite{zhang2020explicit}&$(d-k+h)^{\binom{n}{h}}$&
$|\F|\geq n+d-k$&$dN\cdot\frac{h}{d-k+h}$&Optimal access\\
\hline
The MSCR code in \cite{ye2020new}&$(d-k+h)\cdot(d-k+1)^n$&
$|\F|\geq (d-k+1)n$&$dN$& Full access\\
\hline
New MSCR code&$(d-k+h)\cdot(d-k+1)^n$&$|\F|\geq n+d-k$&
$dN(1-\frac{d-k}{d-k+h}(1-\frac{1}{d-k+1})^h)$&Low access\\
\hline
\end{tabular}
\end{center}
\end{table}

The remainder of this paper is organized as follows. Section
\ref{sec-preliminaries} introduces some necessary preliminaries.
Section \ref{sec_construction} presents the new repair scheme and
its properties.  Section \ref{sec_access} analyzes the access property of
the new repair scheme.  Section
\ref{sec_conclusions} concludes this paper.

\section{Preliminaries}\label{sec-preliminaries}

In this section, we briefly review some necessary  repair properties.
We begin with some notation throughout this paper. Denote
$[a,b)$ and $[a,b]$  respectively the sets $\{a,a+1,\cdots,b-1\}$ and $\{a,a+1,\cdots,b\}$.
We consider the linear vector code that encodes a file
with length $B$ over $\F$ into a codeword with length
$n$, i.e., $\mathbf{C}=(\mathbf{C}_0,\mathbf{C}_1,\ldots,\mathbf{C}_{n-1})$ with $\mathbf{C}_i\in \F^{N},i\in [0,n-1]$ being stored across $n$ nodes, such that any $k$ nodes are capable to recover the original file.  In this paper, we focus on the minimum storage case that the original information is of length $B=kN$, i.e, the MDS case denoted as $(n,k,N)$ MDS code.

\subsection{Three node repair models}

In this subsection, we review three node repair models.

\subsubsection{Single node repair}

In view of the universality of single node failure scenario \cite{rashmi2013solution}, Dimakis \textit{et al.} \cite{dimakis2010network} put forward single node repair model for vector MDS codes. In order to repair one failed node of an $(n,k,N)$ MDS code, we connect any $d$ surviving nodes, called helper nodes, and download $\beta$ symbols from each, where $k\leq d\leq n-1$.
In this model, the repair bandwidth, defined as amount of data downloaded during the repair process, is $\gamma=d\beta$.

\begin{lemma}[\cite{dimakis2010network}]
For an $(n,k,N)$ MDS code in the single node repair model, the repair bandwidth $\gamma$
satisfies
\begin{eqnarray*}
\gamma\geq \frac{dN}{d-k+1}.
\end{eqnarray*}
\end{lemma}

\begin{definition}[\cite{dimakis2010network}]
The $(n,k,N)$ MDS code is said to be a minimum storage regenerating (MSR) code  if
\begin{equation*}\label{MSRbound}
\beta=\frac{N}{d-k+1},\,\gamma=\frac{dN}{d-k+1}.
\end{equation*}
\end{definition}

\subsubsection{Centralized repair of multiple nodes}

In the light of the fact that  the failure of multiple nodes is norm in large-scale storage systems \cite{ford2010availability}, Cadambe \textit{et al.} \cite{cadambe2010centralized} introduced centralized repair model for vector MDS codes. For $(n,k,N)$ MDS code, when
$h\geq 2$ nodes fail, one node in the storage system is responsible for repairing the $h$ failed nodes by downloading $\widetilde{\beta}$ symbols from each of any $d$ helper nodes, where $k\leq d\leq n-h$, which results in the total repair bandwidth  $\gamma=d\widetilde{\beta}$.

\begin{lemma}[\cite{cadambe2010centralized}, \cite{rawat2018centralized-repair}]
For an $(n,k,N)$ MDS code in the centralized repair model, the repair bandwidth  $\gamma$ is lower bounded by
\begin{eqnarray*}
\gamma\geq \frac{dhN}{d-k+h}.
\end{eqnarray*}
\end{lemma}

\begin{definition}[\cite{rawat2018centralized-repair}]
In the  centralized repair model, the $(n,k,N)$ MDS code is called a minimum storage multi-node regenerating (MSMR) code if
\begin{equation*}
\widetilde{\beta}=\frac{hN}{d-k+h},\,\gamma=\frac{dhN}{d-k+h}.
\end{equation*}
\end{definition}

\subsubsection{Cooperative repair of multiple nodes}
 In \cite{shum2013cooperative}, Shum and Hu proposed cooperative repair model for failure of  multiple nodes of vector MDS codes. For an $(n,k,N)$ MDS code $\cC$ over $\F$, when there are  $h$ failed nodes, we need to
connect any $d$ helper nodes to repair them, where $k\leq d\leq n-h$. More specifically, let
 \begin{itemize}
  \item $\cE=\{i_1,i_2,\cdots,i_h\}$ denote the indices of the $h$ failed
  nodes, where $h\geq 2$;
  \item  $\cR$ be a $d$-subset of $[0,n)\setminus \cF$
  that denotes the indices of the $d$ helper nodes.
  \end{itemize}
Under the  cooperative repair model, the repair process  is composed of two phases  \cite{hu2010cooperative,shum2013cooperative}:
\begin{itemize}
\item [1)] \textbf{Download phase.} For any $i_j\in \cE$ and $u\in \mathcal{R}$,  node $i_j$ downloads $\beta_1$ symbols over $\F$ from helper node $u$;
\item [2)] \textbf{Cooperative phase.} For any $i_j\in \cE$ and $i_{j'}\in \cE\setminus\{i_j\}$, node $i_j$ downloads $\beta_2$ symbols over $\F$ from node $i_{j'}$.
\end{itemize}
Therefore, the total repair bandwidth is $\gamma=h(d\beta_1+(h-1)\beta_2)$.

\begin{lemma}[\cite{hu2010cooperative,ye2018cooperative}]\label{lemma_cut_set_bound}
For an $(n,k,N)$ MDS code $\mathcal{C}$ in the cooperative repair model, the  repair bandwidth $\gamma$
is lower bounded by
\begin{eqnarray*}\label{eqn_bound}
\gamma\geq \frac{h(d+h-1)N}{d-k+h}.
\end{eqnarray*}
\end{lemma}

\begin{definition}[\cite{hu2010cooperative}, \cite{ye2018cooperative}]\label{Def_MCSR}
In the  cooperative repair model, an $(n,k,N)$ MDS code $\cC$  is \textit{minimum storage cooperative regenerating (MSCR) code} if
\begin{eqnarray*}\label{eqn_bound}
\beta_1= \beta_2=\frac{N}{d-k+h},\,\gamma= \frac{h(d+h-1)N}{d-k+h}.
\end{eqnarray*}
\end{definition}

 It is proved in \cite{rawat2018centralized-repair} that cooperative repair model is stronger than centralized repair model since the optimality of an MDS code under the former implies its optimality under the latter.

\begin{lemma}[\cite{rawat2018centralized-repair}, Proposition 3]\label{lemma_codes-relation}
Given $n,k,h,d$, any MSCR code is also an MSMR code.
\end{lemma}

Further, the cooperative repair model fits distributed storage systems better due to the distributed pattern. Therefore, we concentrate on the cooperative repair model in this paper. Therefore,  the MSCR code is our focus in what follows.

\subsection{Access property}

During node repair to transmit the target data, the helper node needs to access data from its disk, which will bring disk I/O overhead to the storage system. However, I/Os are a scarce resource in the storage systems:
\begin{itemize}
\item Since the connection speed of the next generation network is accelerating and the storage capacity of a single storage device is increasing, I/O is becoming the main bottleneck of the storage system performance \cite{rashmi2015access}.

\item  Numerous applications of today's storage system services are I/O bounded, for instance, applications that serve a great deal of users to respond \cite{beaver2010haystack} or implement data intensive computing (e.g., analytics \cite{hadoop}).
\end{itemize}
Hence, it is preferred to improve the disk I/O performance in the storage systems, i.e., the total amount of data accessed from disk for repair, denoted by $\gamma_A$,  should be as small as possible.

%For single node repair, assume that to transfer data in the repair process, each helper node accesses $\beta^\ast$ symbols. Then, $\gamma_A=d\beta^\ast$. As each of the $\beta$ symbols that is sent by the helper node is a function of the access data, $\beta^\ast\geq \beta$, which results in $\gamma_A\geq d\beta$. Accordingly, the MSR code is called to be \textit{optimal-access} if $\beta^\ast=\frac{N}{d-k+1},\,\gamma_A=\frac{dN}{d-k+1}$.
%
%For centralized repair of multiple nodes, suppose that  for the sake of sending data during the repair process, each helper node accesses $\widetilde{\beta}^\ast$ symbols. Then, $\gamma_A=d\widetilde{\beta}^\ast$. In the same argument as the single node repair model, we have that $\widetilde{\beta}^\ast\geq \widetilde{\beta}$ and
%$\gamma_A\geq d\widetilde{\beta}$. Accordingly, the MSMR code is called having \textit{optimal-access} property if $\widetilde{\beta}^\ast=\frac{hN}{d-k+h},\,\gamma_A=\frac{dhN}{d-k+h}$.

In cooperative repair model, the data obtained during the download phase is in the memory of the replace nodes firstly and then stored at their disks, so the helper nodes can transfer the data directly from its memory instead of disk in the cooperative phase \cite{zhang2020explicit}. That is,  $\gamma_A$ only
includes the accessed data from helper nodes during the download phase. Precisely, for any $i_j\in \cE$ and $u\in \mathcal{R}$, assume that in order to transmit the data to node $i_j$ during the download phase, the helper node $u$ accesses the amount $T_{i_j,u}$ of its stored data. To avoid repeatedly accessing data, each helper node firstly accesses all the needed data from the disk, and then calculate the data to be transmitted to the failed nodes $\cE=\{i_1,i_2,\cdots,i_h\}$ through these accessed data  \cite{ye2017explicit}. This is to say, the total amount of data accessed from disk for repair is
\begin{eqnarray}\label{eqn_gammaA_Def}
\gamma_A=\sum_{u\in \mathcal{R}}\left|\bigcup_{i_j\in \cE}T_{i_j,u}\right|.
\end{eqnarray}

According to Lemma \ref{lemma_codes-relation}, for an $(n,k,N)$ MSCR code and any given
$u\in \cR$ the amount of data accessed from the helper node $u$ is lower bounded as
$|\bigcup_{i_j\in \cE}T_{i_j,u}|\geq \frac{hN}{d-k+h}$, and then
\begin{eqnarray}\label{eqn_A_bound}
\gamma_A\ge \frac{dhN}{d-k+h}.
\end{eqnarray}
Particularly,  the  MSCR code is said to be  \textit{optimal-access} if the equality in \eqref{eqn_A_bound} is achieved.

\section{A class of MSCR codes}\label{sec_construction}

In this section, we consider the repair scheme for multiple failed nodes
for a known construction in \cite{chen2019explicit} by Chen and Barg.
Although there is a known optimal repair
scheme for a single failed node, the repair scheme for multiple failed nodes is
still open.  To solve the problem, we first slightly modify the original construction. Then we propose an optimal cooperative repair scheme for multiple failed nodes of the code yielded by the modified construction.

\subsection{A construction of MCSR codes}

First of all, we recall the known construction of MSR codes in \cite{chen2019explicit}. Let $n$, $k$, $d$, $r$, $s$, and $\bar{N}$ be positive integers,
where $k=n-r$, $k<d\leq n-1$, $s=d-k+1$, and $\bar{N}=s^n$.
Let $\mathbf{\F}$ be a finite field of size $|\F|\geq n+s-1=n+d-k$, and let
$\lambda_0,\cdots,\lambda_{n-1},\mu_1,\cdots,\mu_{s-1}$ be $n+s-1$ different
elements in $\F$.
For $a\in [0,s^n)$, $\bm a=(a_0,a_1,\cdots,a_{n-1})$ is denoted as the $s$-ary expansion of $a$, i.e., $a=\sum_{i=0}^{n-1}a_is^i$, where $a_i\in [0,s)$ for $i\in [0,n)$. Define
\begin{equation*}
\bm a(i,v)\triangleq(a_0,\cdots,a_{i-1},v,a_{i+1},\cdots,a_{n-1}),
\end{equation*}
that is, $\sum_{j\in [0,n),j\ne i}a_js^j+vs^i$ is the corresponding integer of $\bm a(i,v)$.

Let $\bar{\mathcal{C}}=\{(\bar{\textbf{C}}_0,\bar{\textbf{C}}_1,\cdots,\bar{\textbf{C}}_{n-1})\}$ be an $(n,k=n-r,\bar{N}=s^n)$ vector code,
where each codeword $\bar{\textbf{C}}=(\bar{\textbf{C}}_0,\bar{\textbf{C}}_1,\cdots,\bar{\textbf{C}}_{n-1})$ with
$\bar{\textbf{C}}_i=(\bar{c}_{i,\bm a})_{\bm a\in \Z^n_s}\in \F^{s^n}$ satisfies the following
parity-checking equations over $\F$:
\begin{eqnarray}\label{eqn_cons_chen}
\sum_{i=0}^{n-1}\lambda_i^t\bar{c}_{i,\bm a}+\sum_{i=0}^{n-1}\delta(a_i)\sum_{e=1}^{s-1}
\mu_e^t\bar{c}_{i,\bm a(i,e)}=0,\,\,\textrm{for}\,t\in [0,r-1],\bm a\in \Z^n_{s},
\end{eqnarray}
where the function $\delta(x)$ is defined as
\begin{equation*}
\delta(x)=
\left\{\begin{array}{rl}
1, &\text{if}\,x=0,\\
0, &\text{otherwise}.
\end{array}
\right.
\end{equation*}

\begin{lemma}[\cite{chen2019explicit}]\label{lemma_MDS}
The code $\bar{\mathcal{C}}$ defined by \eqref{eqn_cons_chen} is an  MDS code.
\end{lemma}

In \cite{chen2019explicit}, its repair scheme  for single failed node was proposed. However, the repair scheme for multiple failed nodes is still open. Therefor, in this paper we are going to find an optimal
cooperative repair scheme for multiple failed nodes of code $\bar{\mathcal{C}}$. To this
end, in what follows we slightly modify the original construction.

\begin{construction}\label{cons}
Let $n$, $k$, $d$, $h$, $r$, $s$, and $N$ be positive integers such that
$k=n-r$, $k<d\leq n-1$, $1\leq h\leq n-d$, $s=d-k+1$, and $N=(d-k+h)s^n$.
Let $\cC$ be an $(n,k,N)$ vector code whose codeword $\mathbf{C}=({\mathbf{C}}_0, {\mathbf{C}}_1,\cdots, {\mathbf{C}}_{n-1})\in \cC$
with ${\mathbf{C}}_i=(c_{i,b,\bm a})_{b\in [1,d-k+h],\bm a\in \Z^n_s}$ defined by parity-check equations
\begin{equation}\label{eqn_cons}
\sum_{i=0}^{n-1}\lambda_i^tc_{i,b,\bm a}+\sum_{i=0}^{n-1}\delta(a_i)
\sum_{e=1}^{s-1}\mu_e^tc_{i,b,\bm a(i,e)}=0,\,\,\text{for}\,\,t\in [0,r-1],b\in [1,d-k+h],\bm a\in \Z^n_s.
\end{equation}

\end{construction}

This construction is a slight modified version of the original code $\bar{\mathcal{C}}$,
where the basic idea is to use the space sharing technique, i.e, concatenate several codewords.
For given  $b\in [d-k+h]$, the punctured  one is exactly  a codeword of $\bar{\mathcal{C}}$.
However, this construction
does not contain the original one as a special case since in the modified construction
$d-k+h>1$.

The following lemma follows directly from
Lemma \ref{lemma_MDS}.

\begin{theorem}\label{them_MDS_cons2}
The code $\cC$ generated by Construction \ref{cons} is an MDS code.
\end{theorem}

\subsection{Cooperative repair scheme}

Before presenting cooperative repair scheme for the code generated by Construction \ref{cons}, we introduce a notation and two lemmas.
For $i\in [0,n)$, define
\begin{equation*}
  V_i\triangleq\{{\bm a}=(a_0,a_1,\cdots,a_{n-1})\in \Z^n_{s}\,:\,a_i=0\}.
\end{equation*}

\begin{lemma}\label{lemma_repair}
Let $\cC$ be the code generated by Construction \ref{cons}.
Given $i\in [0,n)$, $b_1\neq b_2\in [1,d-k+h]$ and integer $0<v<s$,
by downloading
\begin{eqnarray*}
c_{u,b_1,\bm a}+c_{u,b_2,\bm a(i,v)}~\textrm{for}~u\in \cR~\textrm{and}~\bm a\in V_{i}
\end{eqnarray*}
node $i$ can recover
\begin{itemize}
  \item $c_{j,b_1,\bm a}+c_{j,b_2,\bm a(i,v)}$ for $j\in [0,n)\setminus \cR$ and $\bm a\in V_{i}$; and
  \item $c_{i,b_1,\bm a(i,e)}$ for $\bm a\in V_i$ and $e\in [1,s-1]$.
\end{itemize}
\end{lemma}

\begin{IEEEproof}
According to \eqref{eqn_cons}, we have
\begin{equation}\label{eqn_repair_v_i-1}
\sum_{j=0}^{n-1}\lambda_j^tc_{j,b,\bm a}+\sum_{e=1}^{s-1}\mu_e^t\sum_{j=0}^{n-1}\delta(a_j)
c_{j,b,\bm a(j,e)}=0, \,\,\text{for }t\in [0,r-1],
\end{equation}
which implies for $\bm a\in V_i$ and $\bm a'\triangleq\bm a(i,v)$
\begin{equation*}
\sum_{j=0}^{n-1}(\lambda_j^tc_{j,b_1,\bm a}+\lambda_j^tc_{j,b_2,\bm a'})+\sum_{e=1}^{s-1}\mu_e^t\sum_{j=0}^{n-1}(\delta(a_j)
c_{j,b_1,\bm a(j,e)}+\delta(a_j')c_{j,b_2,\bm a'(j,e)})=0, \,\,\text{for }t\in [0,r-1],
\end{equation*}
where we apply \eqref{eqn_repair_v_i-1} for $b_1,\bm a$ and $b_2,\bm a'$.
This is to say
\begin{equation}\label{eqn_repair_v_i}
\begin{pmatrix}
\lambda^0_0&\lambda^0_1&\cdots&\lambda^0_{n-1}&\mu^0_0&\mu^0_1&\cdots&\mu^0_{s-1}\\
\lambda^1_0&\lambda^1_1&\cdots&\lambda^1_{n-1}&\mu^1_0&\mu^1_1&\cdots&\mu^1_{s-1}\\
\vdots&\vdots&\ddots&\vdots&\vdots&\vdots&\ddots&\vdots&\\
\lambda^{r-1}_0&\lambda^{r-1}_1&\cdots&\lambda^{r-1}_{n-1}&\mu^{r-1}_0&\mu^{r-1}_1&\cdots&\mu^{r-1}_{s-1}\\
\end{pmatrix}
(\widetilde{c}_{0,\bm a},\widetilde{c}_{1,\bm a},\cdots,\widetilde{c}_{n-1,\bm a},
\Delta_{1,\bm a},\Delta_{2,\bm a},\cdots,\Delta_{s-1,\bm a})^{T}=\bm 0,
\end{equation}
where
\begin{equation*}
\widetilde{c}_{j,\bm a}\triangleq c_{j,b_1,\bm a}+c_{j,b_2,\bm a'}\,\,\text{for }j\in [0,n)
\end{equation*}
and
\begin{equation*}
\Delta_{e,\bm a}\triangleq\sum_{j=0}^{n-1}(\delta(a_j)
c_{j,b_1,\bm a(j,e)}+\delta(a_j')c_{j,b_2,\bm a'(j,e)}) \,\,\text{for}\,\, e\in [1,s-1].
\end{equation*}
Thus, by downloading $\{\widetilde{c}_{j,\bm a}=c_{j,b_1,\bm a}+c_{j,b_2,\bm a'}\,:\,j\in \cR,\, \bm a \in V_i\}$ with $|\cR|=n-r+s-1$,
for any $\bm a\in V_i$ there are $n+s-1-|\cR|=r$ unknown variables in the $r$ equations given by \eqref{eqn_repair_v_i}. Then,
we can recover
\begin{eqnarray}\label{Eqn_known_data}
\{\widetilde{c}_{j,\bm a}=c_{j,b_1,\bm a}+c_{j,b_2,\bm a'}\,:\,j\in [0,n)\setminus \cR,\,\bm a\in V_i\}
\end{eqnarray}
and
\begin{eqnarray*}
\left\{\Delta_{e,\bm a}=\sum_{j=0}^{n-1}(\delta(a_j)
c_{j,b_1,\bm a(j,e)}+\delta(a_j')c_{j,b_2,\bm a'(j,e)})\,:\,e\in [1,s-1],\,\bm a\in V_i\right\}
\end{eqnarray*}
by means of the property of Vandermonde matrix.

Recall that $\bm a\in V_i$ and $\bm a'=\bm a(i,v)$. Thus, for $\bm a\in V_i$, we have $\bm a(j,e)\in V_i$ and $a_j=a'_j$, i.e., $\delta(a_j)=\delta(a'_j)$,
 for any $j\in [0,n)$ with $j\ne i$ and $e\in [1,s-1]$, which means
\begin{equation*}
\begin{split}
\Delta_{e,\bm a}=&\sum_{j=0}^{n-1}(\delta(a_j)
c_{j,b_1,\bm a(j,e)}+\delta(a_j')c_{j,b_2,\bm a'(j,e)})\\
=&\delta(a_i)
c_{i,b_1,\bm a(i,e)}+\delta(a_i')c_{i,b_2,\bm a'(i,e)}+\sum_{j=0,j\ne i}^{n-1}(\delta(a_j)
c_{j,b_1,\bm a(j,e)}+\delta(a_j')c_{j,b_2,\bm a'(j,e)})\\
=&\delta(a_i)
c_{i,b_1,\bm a(i,e)}+\delta(a_i')c_{i,b_2,\bm a'(i,e)}+\sum_{j=0,j\ne i}^{n-1}\delta(a_j)(
c_{j,b_1,\bm a(j,e)}+c_{j,b_2,\bm a'(j,e)})\\
=&\delta(a_i)
c_{i,b_1,\bm a(i,e)}+\delta(a_i')c_{i,b_2,\bm a'(i,e)}+\widetilde{\Delta}_{e,\bm a},
\end{split}
\end{equation*}
where the item $$\widetilde{\Delta}_{e,\bm a}\triangleq \sum_{j=0,j\ne i}^{n-1}\delta(a_j)(
c_{j,b_1,\bm a(j,e)}+c_{j,b_2,\bm a'(j,e)})$$ is a linear combination of recovered data in \eqref{Eqn_known_data} due to $\bm a(j,e)\in V_i$.
Hence, from $\Delta_{e,\bm a}$ we can figure out
$$\delta(a_i)c_{i,b_1,\bm a(i,e)}+\delta(a'_i)c_{i,b_2,\bm a'(i,e)}=c_{i,b_1,\bm a(i,e)}$$
for $e\in [1,s-1]$ and $\bm a\in V_i$,
because of the facts that $\delta(a_i)=1$ and $\delta(a'_i)=0$ for $\bm a\in V_i$.
That is to say, node $i$ can recover
$\{c_{i,b_1,\bm a(i,e)}\,:\,\bm a\in V_i,\, e\in [1,s-1]\}$. This completes the proof.
\end{IEEEproof}

Similarly to Lemma \ref{lemma_repair}, we can prove the following lemma.

\begin{lemma}\label{lemma_repair_one_node}
Let $\cC$ be the code generated by Construction \ref{cons}.  Given
$i\in [0,n)$ and $b\in [d-k+h]$,
by downloading $\{c_{u,b,\bm a}\,:\,u\in \cR,\,\bm a\in V_i\}$ node $i$ can recover
\begin{itemize}
  \item $c_{j,b,\bm a}$ for $j\in [0,n)$  and $\bm a\in V_i$; and
  \item $c_{i,b,\bm a}$ for $\bm a\in \Z^n_s$.
\end{itemize}
\end{lemma}

\vspace{3mm}
Now we are ready to propose a cooperative repair scheme for the code $\cC$ generated by Construction \ref{cons}.

{\bf Download phase}:  For $1\leq j\leq h$, node $i_j\in \cE$ downloads
\begin{equation}\label{eqn_downloading_D1}
D_{1,j}\triangleq\{c_{u,d-k+j,\bm a}\,:\,u\in \cR,\,\bm a\in V_{i_j}\}
\end{equation}
and
\begin{equation}\label{eqn_downloading_D2}
D_{2,j}\triangleq\{c_{u,b,\bm a}+c_{u,d-k+j,\bm a_{(i_j,b)}}\,:\,u\in \cR,\,b\in [1,d-k],\,\bm a\in V_{i_j}\}.
\end{equation}

Firstly,  by Lemma \ref{lemma_repair} node $i_j \,(1\leq j\leq h)$  can recover
\begin{equation}\label{eqn_rec_D_j3}
c_{i_l,b,\bm a}+c_{i_l,d-k+j,\bm a(i_j,b)}\,\,\textrm{for}~i_l\in \cE,\,b\in [1,d-k]~\textrm{and}~\bm a\in V_{i_j}
\end{equation}
and
\begin{equation}\label{eqn_rec_D_j4}
c_{i_j,b,\bm a(i_j,e)}\,\,\textrm{for}~b\in [1,d-k],\,\bm a\in V_{i_j}~\textrm{and}~e\in [1,s-1]
\end{equation}
by downloading $D_{2,j}$. Secondly, according to Lemma \ref{lemma_repair_one_node},   node $i_j \,(1\leq j\leq h)$  can recover
\begin{equation}\label{eqn_rec_D_j1}
c_{i_j,d-k+j,\bm a}\,\,\textrm{for}~\bm a\in \Z^n_{s}
\end{equation}
and
\begin{equation}\label{eqn_rec_D_j2}
c_{i_l,d-k+j,\bm a}\,\,\textrm{for}~i_l\in \cE\setminus\{i_j\}~\textrm{and}~\bm a\in V_{i_j}
\end{equation}
by downloading $D_{1,j}$.

Combining \eqref{eqn_rec_D_j3}-\eqref{eqn_rec_D_j2}, for $1\leq j\leq h$, by downloading $D_{1,j}$ and $D_{2,j}$
node $i_j \,(1\leq j\leq h)$  is able to recover
\begin{equation}\label{eqn_rec_data}
\begin{cases}
c_{i_j,b,\bm a}\,\,&\textrm{for}~b=1,\cdots,d-k,d-k+j~\textrm{and}~\bm a\in \Z^{n}_{s},\\
c_{i_l,d-k+j,\bm a}\,\,&\textrm{for}~i_l\in \cE\setminus\{i_j\}~\textrm{and}
~\bm a\in V_{i_j},\\
c_{i_l,b,\bm a}+c_{i_l,d-k+j,\bm a(i_j,b)}\,\,&\textrm{for}~i_l\in \cE\setminus \{i_j\},b\in [1,d-k]~\textrm{and}~\bm a\in V_{i_j}.
\end{cases}
\end{equation}

{\bf Cooperative phase}: First,  for repairing node $i_j$ for $1\leq j\leq h$,  node $i_l\, (i_l\in \cE\setminus\{i_j\})$
transfers the data
\begin{equation}\label{eqn_downloading_D3}
\begin{cases}
c_{i_j,d-k+l,\bm a}\,\,&\textrm{for}~\bm a\in V_{i_l},\\
c_{i_j,b,\bm a}+c_{i_j,d-k+l,\bm a(i_l,b)}\,\,&\textrm{for}~b\in [1,d-k]~\textrm{and}~\bm a\in V_{i_l}
\end{cases}
\end{equation}
obtained in \eqref{eqn_rec_data} to the target node $i_j$.

Next,  utilizing  the data in \eqref{eqn_rec_data} and \eqref{eqn_downloading_D3}, for $1\leq j\leq h$, node  $i_j$
can recover
\begin{equation}\label{eqn_rec_cor}
c_{i_j,d-k+l,\bm a}\,\,\textrm{for}~l\in [1,h]\setminus\{j\}~\textrm{and}~\bm a\in \Z^n_{s}.
\end{equation}

Finally, based on the data in \eqref{eqn_rec_data} and \eqref{eqn_rec_cor},   node  $i_j$ for $1\leq j\leq h$
can recover all of its data, i.e.,
\begin{equation*}
c_{i_j,b,\bm a}\,\,\textrm{for}~b\in[1,d-k+h]~\textrm{and}~\bm a\in \Z^n_{s}.
\end{equation*}

\begin{theorem}
Let $\cC$ be the code generated by Construction \ref{cons}. Then,
any  $h$ failed nodes  can be recovered by each failed node
\begin{itemize}
  \item downloading $\frac{N}{d-k+h}$ symbols over $\F$ from
  each of $d$ helper node  during the download phase; and
  \item downloading $\frac{N}{d-k+h}$ symbols over $\F$ from each of
  other $h-1$ failed nodes during the cooperative phase,
  \end{itemize}
 which means the code $\cC$ is an MSCR code.
\end{theorem}

\begin{IEEEproof}
According to the preceding cooperative repair scheme, we only need to calculate the amount of data downloaded during the repair process.

During the download phase, in \eqref{eqn_downloading_D1} and \eqref{eqn_downloading_D2}, one downloads
$$(d-k+1)s^{n-1}=\frac{N}{d-k+h}$$
 symbols over $\F$ from each helper node with index in $\cR$, where recall that $s=d-k+1$ and $N=(d-k+h)s^n$.

Similarly, during the cooperative phase, according to \eqref{eqn_downloading_D3}, for any $i_j\in \cE$ the
node $i_j$ downloads
$$(d-k+1)s^{n-1}=\frac{N}{d-k+h}$$
symbols over $\F$ from each node with index in $\cE\setminus \{i_j\}$.
Thus, it follows from Definition \ref{Def_MCSR} that $\cC$ is an MSCR code, which completes the proof.

\end{IEEEproof}

In what follows, we provide an example for the code $\cC$ generated by
Construction \ref{cons} to illustrate its repair scheme.

\begin{Example}\label{example}
Let $n=4$, $r=3$, $k=1$, $d=h=2$, and $s=d-k+1=2$. The $(n,k=n-r,N=(d-k+h)s^n)=(4,1,3\cdot 2^4)$ code $\mathcal{C}$ over $\F$ satisfies the following series of parity-check equations:
\begin{eqnarray*}
&&\sum_{i=0}^3\lambda_i^tc_{i,b,\bm a}+\sum_{i=0}^3\delta(a_i)\mu^tc_{i,b,
\bm a(i,1)}=0,\,t\in [0,2],b\in [1,3],\bm a\in \Z_2^4,
\end{eqnarray*}
where $\lambda_0,\cdots,\lambda_3,\mu$ are the different elements in $\F$ with $|\F|\geq 5$.
Tables \ref{Example download phase} and \ref{Example cooperative phase} show the repair scheme where the $h=2$ failed nodes $\cE=\{0,1\}$ can be recovered by the $d=2$ helper nodes $\mathcal{R}=\{2,3\}$.
Let
\begin{eqnarray*}
V_0=\{\bm a=(a_0,a_1,a_2,a_3)\in \Z_2^4\,:\,a_0=0\}
\end{eqnarray*}
and
\begin{eqnarray*}
V_1=\{\bm a=(a_0,a_1,a_2,a_3)\in \Z_2^4\,:\,a_1=0\}.
\end{eqnarray*}

\begin{table}[htbp]
\begin{center}
\caption{The download phase of the repair scheme}\label{Example download phase}
\begin{tabular}{|c|c|c|}
\hline
Failed node &0&1\\
\hline
Download&$\begin{array}{ll}
\{c_{u,2,\bm a}\,:\,u\in \cR,\,\bm a\in V_0\}\\
\{c_{u,1,\bm a}+c_{u,2,\bm a(0,1)}\,:\,u\in\cR,\,\bm a\in V_0\}\\
\end{array}$&
$\begin{array}{ll}
\{c_{u,3,\bm a}\,:\,u\in \cR,\,\bm a\in V_1\}\\
\{c_{u,1,\bm a}+c_{u,3,\bm a(1,1)}\,:\,u\in\cR,\,\bm a\in V_1\}\\
\end{array}$\\
\hline
Repair&$\begin{array}{ll}
\{c_{0,1,\bm a},c_{0,2,\bm a}\,:\,\bm a\in \Z^4_2\} \\
\{c_{1,1,\bm a}+c_{1,2,\bm a(0,1)},\,c_{1,2,\bm a}\,:\,\bm a\in V_0\}\\
\end{array}$
&$\begin{array}{ll}
\{c_{1,1,\bm a},c_{1,3,\bm a}\,:\,\bm a\in \Z^4_2\} \\
\{c_{0,1,\bm a}+c_{0,3,\bm a(1,1)},\,c_{0,3,\bm a}\,:\,\bm a\in V_1\}\\
\end{array}$\\
\hline
\end{tabular}
\end{center}
\end{table}

\begin{table}[htbp]
\begin{center}
\caption{{The cooperative phase of the repair scheme}}\label{Example cooperative phase}
\begin{tabular}{|c|c|c|}
\hline
Failed node &0&1\\
\hline
Exchange node &1&0\\
\hline
Transfer&$\begin{array}{ll}
\{c_{0,1,\bm a}+c_{0,3,\bm a(1,1)},\,c_{0,3,\bm a}\,:\,\bm a\in V_1\}\\
\end{array}$&$\begin{array}{ll}
\{c_{1,1,\bm a}+c_{1,2,\bm a(0,1)},\,c_{1,2,\bm a}\,:\,\bm a\in V_0\}\\
\end{array}$\\
\hline
Repair&$\begin{array}{ll}
\{c_{0,3,\bm a}\,:\,\bm a\in \Z_2^4\}\\
\end{array}$&
$\begin{array}{ll}
\{c_{1,2,\bm a}\,:\,\bm a\in \Z_2^4\}\\
\end{array}$\\
\hline
\end{tabular}
\end{center}
\end{table}

\end{Example}

\section{Low access property}\label{sec_access}

In this section, we figure out the amount of accessed data at the helper nodes in the download phase for our proposed cooperative repair scheme.
Recall from \eqref{eqn_downloading_D1} and \eqref{eqn_downloading_D2} that  in order to repair the failed nodes $\cE$, for $u\in \mathcal{R}$, the data accessed at the helper node $u$ is
\begin{eqnarray}
&&c_{u,b,\bm a}\,\,\textrm{for}~b\in [d-k+1,d-k+h]~\textrm{and}~\bm a\in \Z_s^n,\label{27}\\
&&c_{u,b,\bm a}\,\,\textrm{for}~b\in [1,d-k]~\textrm{and}~\bm a\in \bigcup_{i_j\in \cE}V_{i_j}.\label{28}
\end{eqnarray}

\begin{theorem}\label{Thm_1903}
For any  $u\in \mathcal{R}$, the amount of the data accessed at the helper node $u$ is $N\cdot G(d-k,h)$,
where
\begin{eqnarray}
G(d-k,h)&\triangleq& 1-\frac{d-k}{d-k+h}\left(1-\frac{1}{d-k+1}\right)^h\label{Eqn_Gd_1}\\
&<&\min\left\{1,\frac{h}{d-k+h}\left(2-\frac{1}{d-k+1}\right)\right\}.\nonumber
\end{eqnarray}
\end{theorem}

\begin{IEEEproof}
The amount of the data in \eqref{27} is
\begin{eqnarray}\label{eqn_access_part1}
h\cdot s^n=h\cdot \frac{N}{d-k+h}.
\end{eqnarray}

Next, we compute the amount of the data in \eqref{28}, which is
$(d-k)\cdot |\bigcup_{i_j\in \cE}V_{i_j}|$. Note that
\begin{eqnarray*}
\Z^{n}_{s}\setminus \bigcup_{i_j\in \cE}V_{i_j}=\{\bm a\,:\,a_{i_j}\ne 0,\,i_j\in\cE,\,\bm a\in \Z^n_s\}
\end{eqnarray*}
with size $s^{n-h}(s-1)^{h}$.
Therefore,
\begin{equation}\label{eqn_access_part2}
\left|\bigcup_{i_j\in \cE}V_{i_j}\right|=s^{n-h}\left(s^h-(s-1)^h\right)=\frac{N}{d-k+h}\left(1-\left(1-\frac{1}{d-k+1}\right)^h\right),
\end{equation}
where $N=(d-k+h)s^n$ and $s=d-k+1$.

Thus, by \eqref{eqn_access_part1} and \eqref{eqn_access_part2}, the amount of the data accessed by the helper node $u\in \mathcal{R}$ is
\begin{eqnarray*}
&&h\cdot \frac{N}{d-k+h}+(d-k)\cdot\frac{N}{d-k+h}\left(1-\left(1-\frac{1}{d-k+1}\right)^h\right)\\
&=&N\cdot\left(1-\frac{d-k}{d-k+h}\left(1-\frac{1}{d-k+1}\right)^h\right),
\end{eqnarray*}
which results in \eqref{Eqn_Gd_1}.

Obviously, $G(d-k,h)<1$.
In addition, we have
\begin{eqnarray}
&&G(d-k,h)\nonumber\\
&<&1-\frac{d-k}{d-k+h}\left(1-\frac{h}{d-k+1}\right)\label{33}\\
&=&\frac{h}{d-k+h}\left(2-\frac{1}{d-k+1}\right),\nonumber
\end{eqnarray}
where \eqref{33} holds due to $(1-\frac{1}{d-k+1})^h>1-\frac{h}{d-k+1}$.
Then, the proof is finished.
\end{IEEEproof}

According to \eqref{eqn_gammaA_Def}, during the repair procedure the total amount of accessed data  is
\begin{eqnarray*}\label{Eqn_GA}
\gamma_A=d N\cdot G(d-k,h).
\end{eqnarray*}
Whereas,  the total amount of access data  for Hadamard MSCR is $dN$ \cite{ye2018cooperative,ye2020new}
and the optimal access amount  is $dN\cdot\frac{h}{d-k+h}$ by \eqref{eqn_A_bound}.  Therefore,
$G(d-k,h)$ can be seen as a ratio of the  amount of  access for the new  MSCR codes over  that for Hadamard MSCR code,
which should be no less than the optimal value ${h\over d-k+h}$.

Note from Theorem \ref{Thm_1903} that: 1) When $d-k\gg h$, then $G(d-k,h)$ tends to be $\frac{h}{d-k+h}(2-\frac{1}{d-k+1})$;
2) For $u\in \mathcal{R}$, $G(d-k,h)$ is strictly less than  2 times of the optimal value $\frac{h}{d-k+h}$.
Some specific comparisons among $G(d-k,h)$, $\frac{h}{d-k+h}(2-\frac{1}{d-k+1})$,
 and $\frac{h}{d-k+h}$ can  be seen in Table \ref{comparison-2}.

\begin{table}[htbp]
\begin{center}
\caption{Some comparisons of the amount of access}\label{comparison-2}
\begin{tabular}{|c|c|c|c|}
\hline
\multirow{2}*{$(d-k,h)$}&\multirow{2}*{$G(d-k,h)$}&\multirow{2}*{$\frac{h}{d-k+h}(2-\frac{1}{d-k+1})$}&(Optimal value)\\
& & &$\frac{h}{d-k+h}$\\
\hline\hline
$(1,2)$&$\approx0.9167$&$1$&$\approx0.6667$\\
\hline
$(2,2)$&$\approx0.7778$&$\approx0.8333$&$0.5$\\
\hline
$(3,2)$&$\approx0.6625$&$0.7$&$0.4$\\
\hline
$(4,2)$&$\approx0.5733$&$0.6$&$\approx0.3333$\\
\hline
$(5,2)$&$\approx0.504$&$\approx0.5238$&$\approx0.2857$\\
\hline
\hline
$(1,3)$&$0.96875$&$1.25$&$0.75$\\
\hline
$(2,3)$&$\approx0.8815$&$1$&$0.6$\\
\hline
$(3,3)$&$\approx0.7891$&$0.875$&$0.5$\\
\hline
$(4,3)$&$\approx0.7074$&$\approx0.7714$&$\approx0.4286$\\
\hline
$(5,3)$&$\approx0.6383$&$0.6875$&$0.375$\\
\hline
\end{tabular}
\end{center}
\end{table}

To demonstrate the  amount of access of the new MSCR code,  we provide an example below.
\begin{Example} Continued with Example \ref{example},  we calculate the amount of data accessed by the $d=2$
helper nodes $2,3$ when
repairing the $2$ failed nodes $0,1$.

Let $V_j=\{\bm a=(a_0,\cdots,a_3)\in \Z_2^4\,:\,a_j=0\}$ for $j\in \cE$, then
we have $|V_0\cup V_1|=2^3+2^3-2^2$. Due to the download phase in Example \ref{example}, to repair the failed nodes $\cE$, for $u\in \mathcal{R}$, the data accessed by node $u$ is
\begin{eqnarray}\label{15}
\left\{\begin{array}{ll}
c_{u,b,\bm a}\,&\textrm{for}~2\leq b\leq 3,\bm a\in \Z_2^4,\\
c_{u,1,\bm a}\,&\textrm{for}~\bm a\in V_0\cup V_1.
\end{array}
\right.
\end{eqnarray}
Accordingly, for $u\in \mathcal{R}$,  the amount of data accessed at node $u$ is
\begin{eqnarray*}
2\cdot 2^4+\left(2^3+2^3-2^2\right)=44
\end{eqnarray*}
by \eqref{15}, and
\begin{eqnarray*}
G(d-k,h)=G(1,2)=\frac{44}{48}\approx0.9167.
\end{eqnarray*}
\end{Example}
\section{Conclusions}\label{sec_conclusions}
In this paper, a modified construction of MDS codes is proposed. Based on this construction,
  an optimal repair scheme for multiple failures was introduced under the cooperative repair model.
  For this repair scheme, specifically, the amount of access is strictly less than  $2$ times the optimal access amount.


\begin{thebibliography}{11}

\bibitem{balaji2018overview} S.B. Balaji, M.N. Krishnan, M. Vajha, V. Ramkumar, B. Sasidharan, and P.V. Kumar, ``Erasure coding for distributed storage: An overview," \textit{Sci. China Inf. Sci.}, vol. 61, Art. no. 100301, Oct. 2018.

\bibitem{beaver2010haystack} D. Beaver, S. Kumar, H.C. Li, J. Sobel, and P. Vajgel, ``Finding a needle in haystack: Facebook's photo storage," in \textit{Proc. of USENIX OSDI}, 2010.

\bibitem{cadambe2010centralized} V.R. Cadambe, S.A. Jafar, H. Maleki, K. Ramchandran, and C. Suh, ``Asymptotic interference alignment for optimal repair of MDS codes in distributed storage,"  \textit{IEEE Trans. Inf. Theory}, vol. 59, no. 5, pp. 2974-2987, May 2013.

\bibitem{chen2019explicit} Z. Chen and A. Barg, ``Explicit constructions of MSR codes for clustered distributed storage: The rack-aware
storage model," \textit{IEEE Trans. Inf. Theory}, vol. 66, no. 2, pp. 886-899, Feb. 2020.

\bibitem{dimakis2010network} A.G. Dimakis, P.B. Godfrey, Y. Wu, M.J. Wainwright, and K. Ramchandran, ``Network coding for distributed storage systems,"  \textit{IEEE Trans. Inf. Theory}, vol. 56, no. 9, pp. 4539-4551, Sep. 2010.

\bibitem{dimakis2011survey} A.G. Dimakis, K. Ramchandran, Y. Wu, and C. Suh, ``A survey on network codes for distributed storage," \textit{Proc. of the IEEE}, vol. 99, no. 3, pp. 476-489, Mar. 2011.

\bibitem{ford2010availability} D. Ford, F. Labelle, F. Popovici, M. Stokely, V. Truong, L. Barroso, C. Grimes, and S. Quinlan, ``Availability in globally distributed storage systems,"  \textit{Proc. of USENIX OSDI},  2010.

\bibitem{hadoop} Hadoop, in \textit{http://hadoop.apache.org}.

\bibitem{han2015update} Y.S. Han, H. Pai, R. Zheng, and W.H. Mow, ``Efficient exact regenerating codes for Byzantine fault tolerance in distributed networked storage,"  \textit{IEEE Trans. Commun.}, vol. 62, no. 2, pp. 385-397, Feb. 2014.

\bibitem{Han15update} Y.S. Han, H. Pai, R. Zheng and P.K. Varshney, ``Update-efficient error-correcting product-matrix codes,"  \textit{IEEE Trans. Commun.}, vol. 63, no. 6, pp. 1925-1938, Jun. 2015.

\bibitem{hu2010cooperative} Y. Hu, Y. Xu, X. Wang, C. Zhan, and P. Li, ``Cooperative recovery of distributed storage systems from multiple losses with network coding,"  \textit{IEEE J. Sel. Areas Commun.}, vol. 28, no. 2, pp. 268-276, Feb. 2010.

\bibitem{Li15Frame} J. Li, X. Tang, and U. Parampalli, ``A framework of constructions of minimal storage regenerating codes
with the optimal access/update property," \textit{IEEE Trans. Inform. Theory,} vol. 61, no. 4, pp. 1920-1932, Apr. 2015.

\bibitem{li2018generic} J. Li, X. Tang, and C. Tian, ``A generic transformation to enable optimal repair in MDS codes for distributed storage systems,"  \textit{IEEE Trans. Inf. Theory}, vol. 64, no. 9, pp. 6257-6267, Sep. 2018.

\bibitem{rashmi2011optimal} K.V. Rashmi, N.B. Shah, and P.V. Kumar, ``Optimal exact-regenerating codes for distributed storage at the MSR and MBR points via a product-matrix construction,"  \textit{IEEE Trans. Inf. Theory}, vol. 57, no. 8, pp. 5227-5239, Aug. 2011.

\bibitem{rashmi2013solution} K.V. Rashmi, N.B. Shah, D. Gu, H. Kuang, D. Borthakur, and K. Ramchandran, ``A solution to the network challenges of data recovery in erasure-coded distributed storage systems: A study on the Facebook warehouse cluster," in \textit{Proc. USENIX HotStorage}, Jun. 2013.

\bibitem{rashmi2015access} K.V. Rashmi, P. Nakkiran, J. Wang, N.B. Shah, and K. Ramchandran, ``Having your cake and eating it too: Jointly optimal erasure codes for I/O, storage, and network-bandwidth," in \textit{Proc. of USENIX FAST}, 2015.

\bibitem{rawat2018centralized-repair} A.S. Rawat, O.O. Koyluoglu, and S. Vishwanath, ``Centralized repair of multiple node failures with applications to communication efficient secret sharing," \textit{IEEE Trans. Inf. Theory}, vol. 64, no. 12, pp. 7529-7550, Dec. 2018.

 \bibitem{shum2013cooperative} K.W. Shum and Y. Hu ``Cooperative regenerating codes," \textit{IEEE Trans. Inf. Theory}, vol. 59, no. 11, pp. 7229-7258, Nov. 2013.

 \bibitem{tamo2012zigzag} I. Tamo, Z. Wang, and J. Bruck, ``Zigzag codes: MDS array codes with optimal rebuilding,"  \textit{IEEE Trans. Inf. Theory}, vol. 59, no. 3, pp. 1597-1616, May 2013.

 \bibitem{vajha2018access} M. Vajha, V. Ramkumar, B. Puranik, G Kini, E. Lobo, B. Sasidharan, P.V. Kumar, A. Barg, Min Ye, S. Narayanamurthy, S. Hussain, and Siddhartha Nandi, ``Clay codes: Moulding MDS codes to yield an MSR code," \textit{Proc. 16th USENIX Conf. File Storage Technol. (FAST)}, Feb. 2018, pp. 139-154.

 \bibitem{Wang11long} Z. Wang, I. Tamo, and J. Bruck, ``On codes for optimal rebuilding access," in \textit{Proc. 49th Annu. Allerton Conf. Commun.,
Control, Comput.,} Monticello, IL,  Sep. 2011, pp. 1374-1381.

\bibitem{wang2016access} Z. Wang, I. Tamo, and J. Bruck, ``Explicit minimum storage regenerating codes,"  \textit{IEEE Trans. Inf. Theory}, vol. 62, no. 8, pp. 4466-4480, Aug. 2016.

\bibitem{ye2017explicit} M. Ye and A. Barg, ``Explicit constructions of high-rate MDS array codes with optimal repair bandwidth,"  \textit{IEEE Trans. Inf. Theory}, vol. 63, no. 4, pp. 2001-2014, Oct. 2017.

\bibitem{ye2018cooperative}  M. Ye and A. Barg, ``Cooperative repair: Constructions of optimal MDS codes for all admissible parameters," \textit{IEEE Trans. Inf. Theory}, vol. 65, no. 3, pp. 1639-1656, Mar. 2019.

\bibitem{ye2020new} M. Ye, ``New constructions of cooperative MSR codes: Reducing node size to $\exp(O(n))$,"  \textit{IEEE Trans. Inf. Theory}, vol. 66, no. 12, pp. 7457-7464, Dec. 2020.

\bibitem{zhang2019scalar}  Y. Zhang and Z. Zhang, ``Scalar MSCR codes via the product matrix construction,"  \textit{IEEE Trans. Inf. Theory}, vol. 66, no. 2, pp. 995-1006, Feb. 2020.

\bibitem{zhang2020explicit}  Y. Zhang, Z. Zhang, and L. Wang, ``Explicit constructions of optimal-access MSCR codes for all parameters,"  \textit{IEEE Communications Letters}, vol. 24, no. 5, pp. 941-945, May 2020.

\end{thebibliography}
\end{document}